\documentstyle[twocolumn]{jpsj2}

\newcommand{\simle}
{\raisebox{-0.75ex}[-1.5ex]{$\;\stackrel{<}{\sim}\;$}}

\def\kz{k_z}
\def\kx{k_x}
\def\ky{k_y}
\def\tz{t_z}

\title{
First-Principles Study of Electronic Structure in
$\alpha$-(BEDT-TTF)$_2$I$_3$ \\ at Ambient Pressure and with Uniaxial Strain. 
}

\author{
Hiori \textsc{Kino} and Tsuyoshi \textsc{Miyazaki}
}
\def\runauthor{Hiori \textsc{Kino} and Tsuyoshi \textsc{Miyazaki}}

\inst{
National Institute for Materials Science, 1-2-1 Sengen, Tsukuba, Ibaraki 305-0047, Japan.
}

\abst{ 
Within the framework of the density functional theory,
we calculate the electronic structure of $\alpha$-(BEDT-TTF)$_2$I$_3$ at 8~K and room temperature at ambient pressure and with uniaxial strain along the $a$- and $b$-axes.
We confirm the existence of anisotropic Dirac cone dispersion near the chemical potential.
We also extract the orthogonal tight-binding parameters to analyze physical properties.
An investigation of the electronic structure near the chemical potential clarifies
that effects of 
uniaxial strain along the $a$-axis is different from that along the $b$-axis. 
The carrier densities show $T^2$ dependence at low temperatures, which
 may explain the experimental findings not only qualitatively  but also quantitatively.
}

\kword{ $\alpha$-(BEDT-TTF)$_2$I$_3$, first-principles study, Dirac cone dispersion, tight-binding model, carrier density}

\begin{document}
\maketitle

\section{Introduction}
A quasi-two-dimensional organic charge transfer conductor, $\alpha$-(BEDT-TTF)$_2$I$_3$, has attracted much attention due to its exotic physical properties. 
The temperature dependence of the resistivity shows a metal-insulator transition at 135~K, which is induced by charge-ordering at low temperatures. \cite{REF_KAJITA_MOBILITY,REF_TAJIMA_SUPER,REF_KONDO_STRAIN}
When hydrostatic pressure or uniaxial strain suppresses the charge-ordering, the temperature dependence of the resistivity shows a metallic behavior even at low temperatures. However, the transport property in this metallic state is very unique. 
It is reported that the carrier density decreases and the mobility
increases drastically with decreasing temperature; 
this results in temperature-independent resistivity. \cite{REF_KAJITA_MOBILITY}
Kajita et al. named this state the narrow-gap semiconducting state.

Applying uniaxial strain is now one of the important techniques of changing the physical properties of organic charge transfer salts. \cite{REF_KAGOSHIMA_REV} 
Tajima et al. reported that the transport property depends on the direction of uniaxial strain. They discovered superconductivity with 1-5 kbar of uniaxial strain along the $a$-axis. \cite{REF_TAJIMA_SUPER}
However, no superconductivity appears with uniaxial strain along the $b$-axis. In addition to this, the metal-insulator transition temperature decreases when uniaxial strain is applied along the $a$-axis, 
while the metal-insulator transition temperature only nominally changes with uniaxial strain along the $b$-axis. 
Thus the obtained phase diagrams are different depending on the directions of uniaxial strain. Kondo and Kagoshima measured the change of the crystal structures with uniaxial strain at 8~K and room temperature (RT) to discuss the change of electronic structures in the semi-empirical estimations of transfer and overlap integrals \cite{REF_KONDO_STRAIN,  REF_KONDO_ATOMIC_PRIVATE, REF_HS, REF_HS_alphaET}. 

Based on these electronic structures in the semi-empirical
approximation, Kobayashi et al. discussed the charge-ordering,
narrow-gap semiconducting and superconducting
states. \cite{REF_KOBAYASHI_THEO_SUPER,REF_KOBOYASHI_THEO_ANOTHER_SUPER}
Through a careful investigation of the electronic structure, Katayama et
al.  also found a unique electronic structure near the Fermi level, called two-dimensional anisotropic Dirac cone dispersion, $E=\pm |{\vec v_F}\cdot {\vec k_{\perp}}|$, where $\vec v_F$ is the Fermi velocity and $\vec k_{\perp}$ is the wave number along the $a^*b^*$-plane measured from the crossing point. \cite{REF_KATAYAMA_DIRAC}. 
This unique feature may be the origin of the narrow-gap semiconducting state, because it is qualitatively consistent with the low carrier density and compensating high mobility. It also leads us to expect large diamagnetism analogous to those observed in graphite and bismuth. 
Although this discovery is very important in clarifying the interesting physical properties of the title compound, 
there is a problem that their studies rely on the simple semi-empirical approximation. 
Therefore, it is important to confirm that the simplicity of their model is not the origin of the interesting electronic structure.

BEDT-TTF charge transfer salts consist of the BEDT-TTF conducting and anion insulating layers.  
The highest occupied molecular orbitals (HOMO) of the BEDT-TTF molecule usually form the electronic structure near the Fermi level. 
The semi-empirical approximation, which evaluates the transfer and
overlap integrals between the HOMOs of the BEDT-TTF molecules, has
achieved great success in the prediction of the electronic structure of the BEDT-TTF charge transfer organic salts at ambient pressure.
One of the reasons behind this success is that the BEDT-TTF charge transfer organic salts usually have large round Fermi surfaces at half-filling, which conceals small errors of the estimated transfer and overlap integrals in the extended H\"uckel approximation.

In contrast to the above, 
 the electronic structure of the title compound is very delicate; it is semimetallic with a very small band overlap (at least in the semi-empirical approximation). 
Though parameters in the semi-empirical approximation are tuned up to reconstruct the Fermi surface of some of BEDT-TTF compounds, a small difference in the electronic structure can change the Fermi surfaces drastically in the title compound because of the semimetallicity. 
Furthermore, the electronic structure at high temperatures may be different from that at low temperatures because the thermal contraction reported from the X-ray measurements is appreciable.
Therefore, an accurate first-principles calculation is demanded for elucidating the reliable electronic structure. 
In this work, we calculate electronic structures for the unit cells at
8~K and at RT, which are determined by the X-ray measurements, to
clarify the change of their electronic structures within the framework
of the density functional theory and the generalized gradient approximation (GGA). In addition, we extract the transfer integrals of the tight-binding model to evaluate carrier densities and other physical properties.

We have demonstrated that the GGA has sufficient accuracy to determine the structures of $\beta$'-(BEDT-TTF)$_2$X (X=ICl$_2$ and AuCl$_2$) at ambient pressure and various applied pressure.\cite{REF_MIYAZAKI_BEDT, REF_MIYAZAKI_BEDT2}. 
Encouraged by these successes, we use the GGA also in this study to relax the internal coordinates and to calculate the electronic structures of the title compound at ambient pressure, with uniaxial strain of 2~kbar along the $a$-axis and  3~kbar along the $b$-axis. 
We employ the ultrasoft pseudopotential technique with planewave basis sets, and the cutoff energies of 25 Ry for the wavefunction and 256 Ry
for the charge density. For $k$-point sampling, we use a 4$\times$4$\times$2 mesh during the relaxation of the internal coordinates and an 8$\times$8$\times$4 mesh to calculate electronic structures in the relaxed atomic positions. In this calculation, we use lattice parameters from the X-ray measurements obtained at 8~K and RT, because we want to study the thermal effects on the electronic structure through the difference in the lattice parameters, which is very large in this material.\cite{Xraysymmetry} For example, the lattice parameters $a$ and $b$ change by 0.5\% with 2~kbar strain in the $a$-axis and by 1\% with 3~kbar strain in the $b$-axis. However, $a$ and $b$ change about 2\% with decreasing temperature (from RT to 8~K). The purpose of this study is to evaluate and discuss changes in electronic structures depending on temperature and uniaxial strain. 

\section{Results and Discussion}

\subsection{Results}

\subsubsection{Overall electronic structure}




The relaxed internal atomic positions are only slightly different from
those determined by the X-ray measurement. The largest difference arises
for atoms near the terminal hydrogens. In the case of ambient pressure at RT and 8~K, the hydrogens move by 0.11~\AA\ at most. 
This may be related to the well-known fact that 
the X-ray measurements underestimate C-H bond lengths.
The other atoms move much less than the hydrogens. Carbon, sulfur and iodine atoms move by 0.04, 0.03 and 0.03~\AA\ at most, respectively.  

\begin{figure}
\begin{center}
\includegraphics[width=8cm]{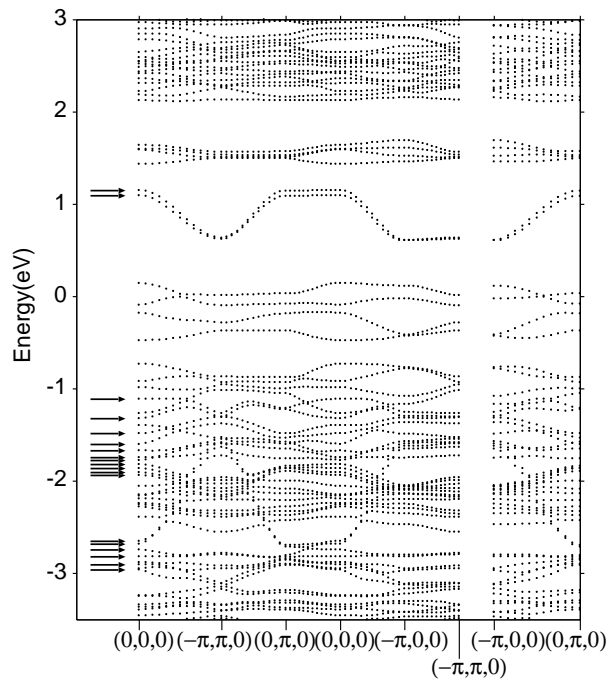}
\end{center}
\caption{ Electronic structure at ambient pressure and 8~K.
Arrows show  the states that have large contributions from 2I$_3$  at the $\Gamma$-point, (0,0,0).
We take the Fermi level to be at $E=0$.
}
\label{dispersionWide}
\end{figure}

We show the electronic structure from -3.5~eV to 3.0~eV in Fig.~\ref{dispersionWide}.
We take the Fermi level to be at $E=0$.
The HOMO-2, HOMO-1, HOMO and LUMO bands in the crystal, which are at -0.5~eV to 0.2eV, are composed of the HOMOs of the BEDT-TTF molecules.
The LUMO+1 and LUMO+2 around +1~eV are composed of I$_3^-$.  There are
14 (7$\times$ 2) molecular levels made of two I$_3^-$ below the Fermi
level in this energy range. The energy bands where the contributions of
I$_3^-$ are large are shown by arrows left of the $\Gamma$ point,
(0,0,0), in Fig.~1.
There are more than 14 arrows because mixing occurs between the BEDT-TTF and I$_3^-$ molecules. 
The bands coming from I$_3^-$ have large dispersion along the $a^*$-axis
because I$_3^-$ molecules form a one-dimensional chain structure along the $a$-axis.

\begin{figure}[p]
\includegraphics[width=17cm]{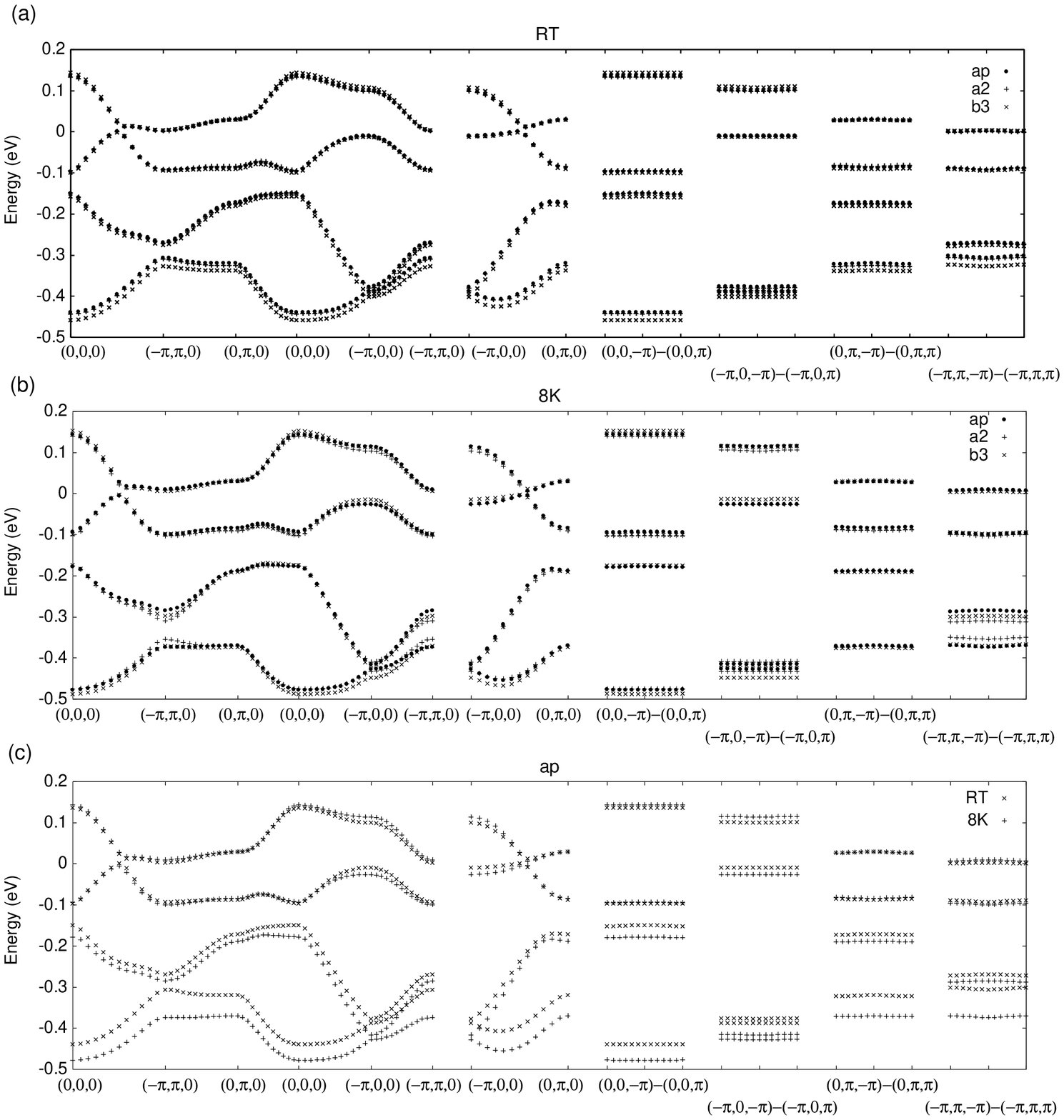}

\caption{
  Electronic structures at ambient pressure (ap,$\bullet$)
   and with uniaxial strain of 2~kbar (a2,$+$) along the $a$-axis
    and  3~kbar along the $b$-axis (b3,$\times$)
    at RT (a) and at 8~K (b).
     A comparison of RT ($\times$) and 8~K ($+$)  at ambient pressure is shown in (c).
We take $E=0$ for the chemical potential. 
     }
     \label{dispersionAll}

     \end{figure}


Let us examine the electronic structure near the Fermi level.
 Figures~\ref{dispersionAll} (a) and \ref{dispersionAll} (b) show the electronic structure at 8~K and RT. 
 The left side shows the electronic structure on the $a^*b^*$-plane and
 the right side shows that along the $c^*$-axis.
 The dispersion along the $c^*$-axis is much smaller than those along the other directions, reflecting the two-dimensional nature of the compound.

 Here we compare the bandwidths at ambient pressure and with strain.
In this paper we define the total bandwidth as the highest energy of the LUMO band minus the lowest energy of the HOMO-2 band, because they are made of the HOMOs of the BEDT-TTF molecules. 
The total bandwidth is slightly larger with strain of 3~kbar along the
$b$-axis than in the other cases;
however, the differences between the $a$- and $b$-axis strain are slight
in the HOMO and LUMO bands at this energy
scale. Figure~\ref{dispersionAll} (c) shows the electronic structures at
ambient pressure for 8~K and RT. The changes according to the
temperature are much larger than those according to uniaxial strain at the same temperature. Details are summarized in Table~\ref{bandwidth}.

Strangely, the bandwidths with strain of 2~kbar along the $a$-axis are
almost the same or slightly smaller than those at ambient pressure at
both 8~K and RT. The bandwidths with uniaxial strain of 3~kbar along the
$b$-axis are larger than those at ambient pressure. 
The bandwidths of HOMO and LUMO are about 0.1 and 0.13$\sim$0.15~eV, respectively, at both 8~K and RT.
 
\begin{table}
\caption{
 Bandwidths (eV) calculated in 8$\times$8$\times$4 mesh
 at ambient pressure (ap), with uniaxial strain of 2~kbar along the $a$-axis (a2) 
         and 3~kbar along the $b$-axis (b3)
	  at RT (a) and 8~K (b).
 H-L represents the highest energy of the LUMO band minus the lowest energy of the HOMO band.
 Total means the highest energy of the LUMO band minus the lowest energy of the HOMO-2 band.
  The state with * is metallic in 8$\times$8$\times$4  mesh. 
}
\begin{tabular}{cccc}
$T=$RT \\ \hline
 & ap & a2* & b3 \\ \hline

 HOMO  & 0.098 & 0.095  & 0.099 \\

 LUMO & 0.136 & 0.134  & 0.141 \\

 H-L & 0.233 & 0.228  & 0.243 \\

 total & 0.576 & 0.576 & 0.602  \\
 \hline
 \end{tabular}

\begin{tabular}{cccc}
$T=$8~K \\ \hline

 & ap & a2 & b3 \\ \hline

HOMO & 0.093 & 0.099  & 0.099 \\

LUMO & 0.141 & 0.137  & 0.151 \\

H-L & 0.243 & 0.243 & 0.253 \\

total & 0.621 & 0.616  & 0.640 \\ \hline
\end{tabular}

\label{bandwidth}
\end{table}

\subsubsection{Anisotropic Dirac cone dispersion}

Next 
we concentrate on the 2D electronic structure to investigate the existence of the anisotropic Dirac cone dispersion discovered by Kobayashi et al.
 through the semi-empirical electronic structure calculation.
We discuss the dispersion along the interlayer direction later in \S.~\ref{sec_interlayer}.

\begin{figure}
\begin{center}
\includegraphics[width=8cm]{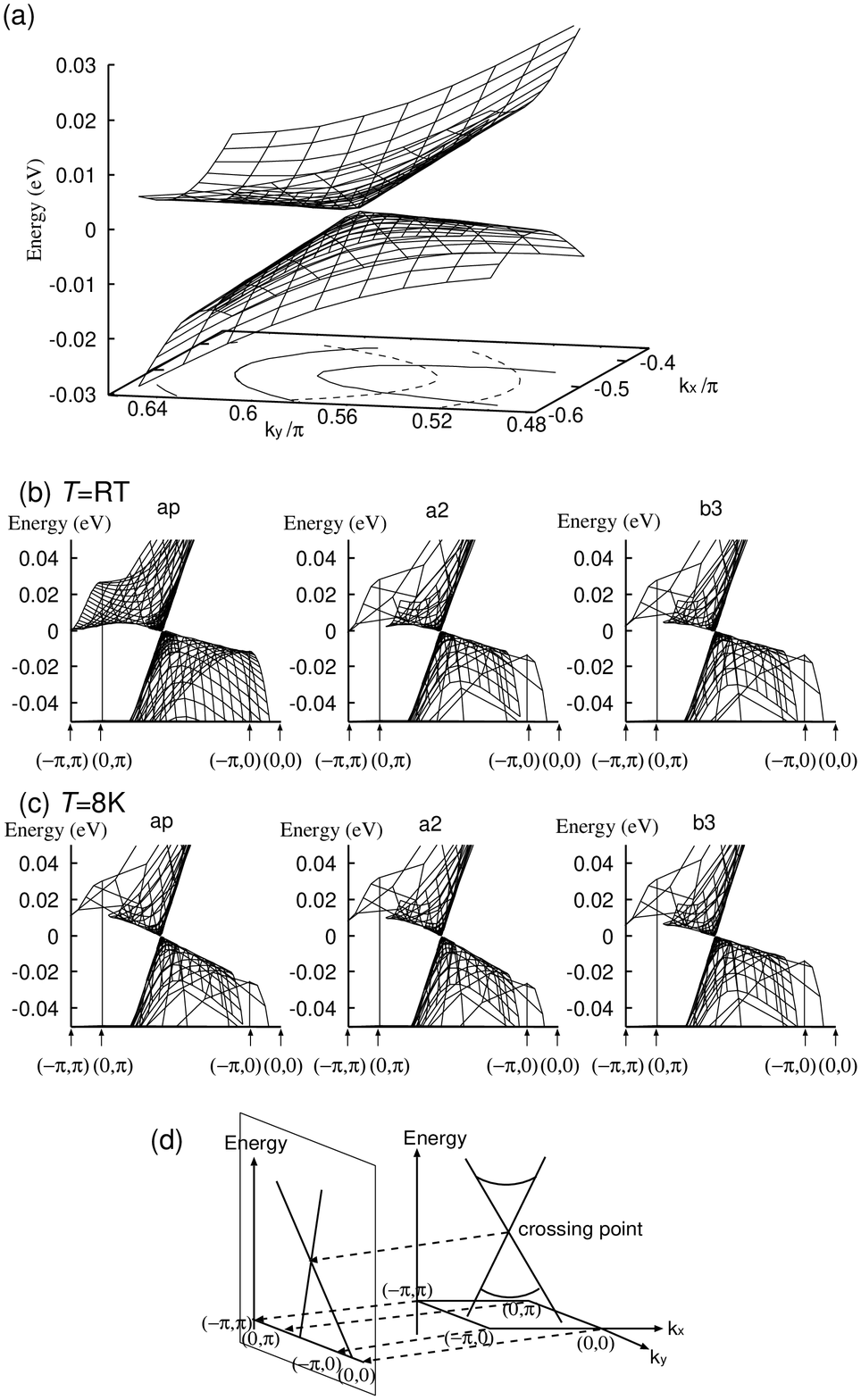}
\end{center}

\caption{
(a) Electronic structure at $\kz=0$ near anisotropic Dirac cone dispersion at ambient pressure and RT.
 Contour plot is also shown at the bottom. Solid (dashed) lines are for the HOMO (LUMO) band.
 $|v_{F,y}|$ is much smaller for $\ky<k_{y0}$ than  for $\ky>k_{y0}$ in the HOMO band,
  while  $|v_{F,y}|$ is much smaller for $\ky>k_{y0}$ than  for $\ky<k_{y0}$ in the LUMO band,
  where  $k_{y0}$ is the wavenumber along the $b^*$-axis at the crossing point of the anisotropic Dirac cone.
  Projections of $E(\kx,\ky,\kz=0)$ 
  in the area from $(-\pi,0)$ to $(0,\pi)$
   onto the plane perpendicular to the $ab$-plane
 at ambient pressure (ap),
  with uniaxial strain of 2~kbar along the $a$-axis (a2)
 and  3~kbar along the $b$-axis (b3) (b) at room temperature, and (c) at 8~K . 
  Vertical lines at (0,$\pi$) and ($-\pi$,0)  are guides for the eye.
  (d) Schematic representation of projection of  $E(\kx,\ky,0)$ in (b) and (c).
We do not remove hidden surfaces in the figures.
    The finest mesh is 324$\times$324 in all cases.
We take $E=0$ to be the crossing energy of the anisotropic Dirac cone dispersion.
    }
    \label{masslessDiracCone}
    \end{figure}

The dispersions of the HOMO and the LUMO from $(-\pi,0)$ to $(0,\pi)$
appear to cross. We examine this region in detail at $\kz=0$, as shown in Fig.~\ref{masslessDiracCone}. 
Figure~\ref{masslessDiracCone} (a) shows the  electronic structure near $(-\pi/2,\pi/2)$ at ambient pressure and RT.
It clearly shows the existence of the anisotropic Dirac cone dispersion,
since the finer we make the mesh of the $k$-point, the smaller the
energy gap becomes. The value of the direct gap is 0.54~meV in the
finest mesh, which corresponds to the 324 $\times$ 324 mesh in the $a^*$- and $b^*$-axis. 
These results strongly support that this system has anisotropic Dirac cone dispersion (accidental degeneracy) in the  $a^*b^*$-plane. \cite{REF_HERRING} 
Note that the symmetry of this system is P$\bar{1}$ and that this
accidental degeneracy occurs not at a high-symmetry point, but at a general point $(\kx,\ky)=(-0.494 \pi, 0.574 \pi)$.
It is known that such kinds of degeneracy at the Fermi level take place, for example, in graphite sheets and carbon nanotubes at a high-symmetry point.
However, the present system is unique because the accidental degeneracy occurs 
{\it  at a general $k$-point } as well as at the Fermi level.

In Figs.~\ref{masslessDiracCone}~(b) and \ref{masslessDiracCone}~(c),
we project the electronic structure in the area from $(\kx,\ky)$=$(-\pi,0)$ to $(0,\pi)$ onto the plane perpendicular to the $a^*b^*$-plane.
Figure~\ref{masslessDiracCone}~(b) shows the electronic structures at RT, while Fig.~\ref{masslessDiracCone}~(c)  shows those at 8~K. 
Figure~\ref{masslessDiracCone}~(d) schematically represents the projection.
These figures clearly show the existence of the anisotropic Dirac cone dispersion in all  cases. At ambient pressure and RT, the energy at $(-\pi,\pi)$ ($E(-\pi,\pi)$) is almost the same  as that of the crossing point of the anisotropic Dirac cone dispersion ($E({\rm cone})$). 
With strain of 2~kbar along the $a$-axis at RT, $E(-\pi,\pi)$ is lower than $E({\rm cone})$.
 On the other hand, with strain of 3~kbar along the $b$-axis at RT, 
 $E(-\pi,\pi)$ is higher than $E({\rm cone})$.
 $E(-\pi,\pi)$ is much higher than $E({\rm cone})$ at 8~K. 
 $E({\rm cone})$ becomes the Fermi level when it is lower than  $E(-\pi,\pi)$ if one  considers only the plane at $\kz=0$.
Details are summarized in Table~\ref{Ediff}. 

There may be errors in the estimated energy. However, it is safe to conclude that there is a tendency that $E(-\pi,\pi)-E({\rm cone})$ at 8~K is higher than that at RT, and that the Dirac cone dispersion determines the transport properties at low temperatures.

\begin{table}

\caption{
Energy differences (meV) between $E(-\pi,\pi)$ and $E({\rm cone})$ at room temperature (RT) 
 and 8~K.
 The van Hove  singularity at $(-\pi,\pi)$ is the nearest the Fermi energy. 
$E({\rm cone})$ is the energy of the crossing point at the Dirac cone dispersion.
}

\begin{tabular}{lccc} $T=$RT \\\hline
&ap & a2 & b3 \\ \hline
$E(-\pi,\pi)-E({\rm cone})$ &
0.8 & -0.5   & 2.0 \\ \hline
\end{tabular}

\begin{tabular}{lccc} $T=$8~K \\\hline
& ap & a2 & b3 \\ \hline
$E(-\pi,\pi)-E({\rm cone})$ &
11.1 & 8.6  & 6.1 \\ \hline
\end{tabular}
\label{Ediff}
\end{table}

\subsection{Discussion}
 
\subsubsection{Bandwidths}
The calculated results show that the total band width with uniaxial strain of 3~kbar along the $b$-axis is much larger than those at ambient pressure and with uniaxial strain of 2~kbar along the $a$-axis. 
Taking into account the fact that the value of strain along the $b$-axis
is 50\% larger than that along the $a$-axis, the change of the total
bandwidth with  strain along the $b$-axis is much larger (more than
150\%) than that with strain along the $a$-axis. 
Therefore, we can expect that the bandwidth with strain along the
$b$-axis is larger than that with strain along the $a$-axis if the values of strain are the same.
The charge-ordered phase is usually discussed in terms of  transfer
integrals and long-range Coulomb interaction. Roughly speaking, the case
where the bandwidth is comparable to the Coulomb interaction
determines the phase boundary. Thus, it is expected that uniaxial strain
along the $b$-axis destroys the CO phase slightly more easily than that
with strain along the $a$-axis. This agrees with the experimental findings.


\subsubsection{Analysis using tight-binding model for the $ab$-plane}

\begin{figure}
\begin{center}
\includegraphics[width=7cm]{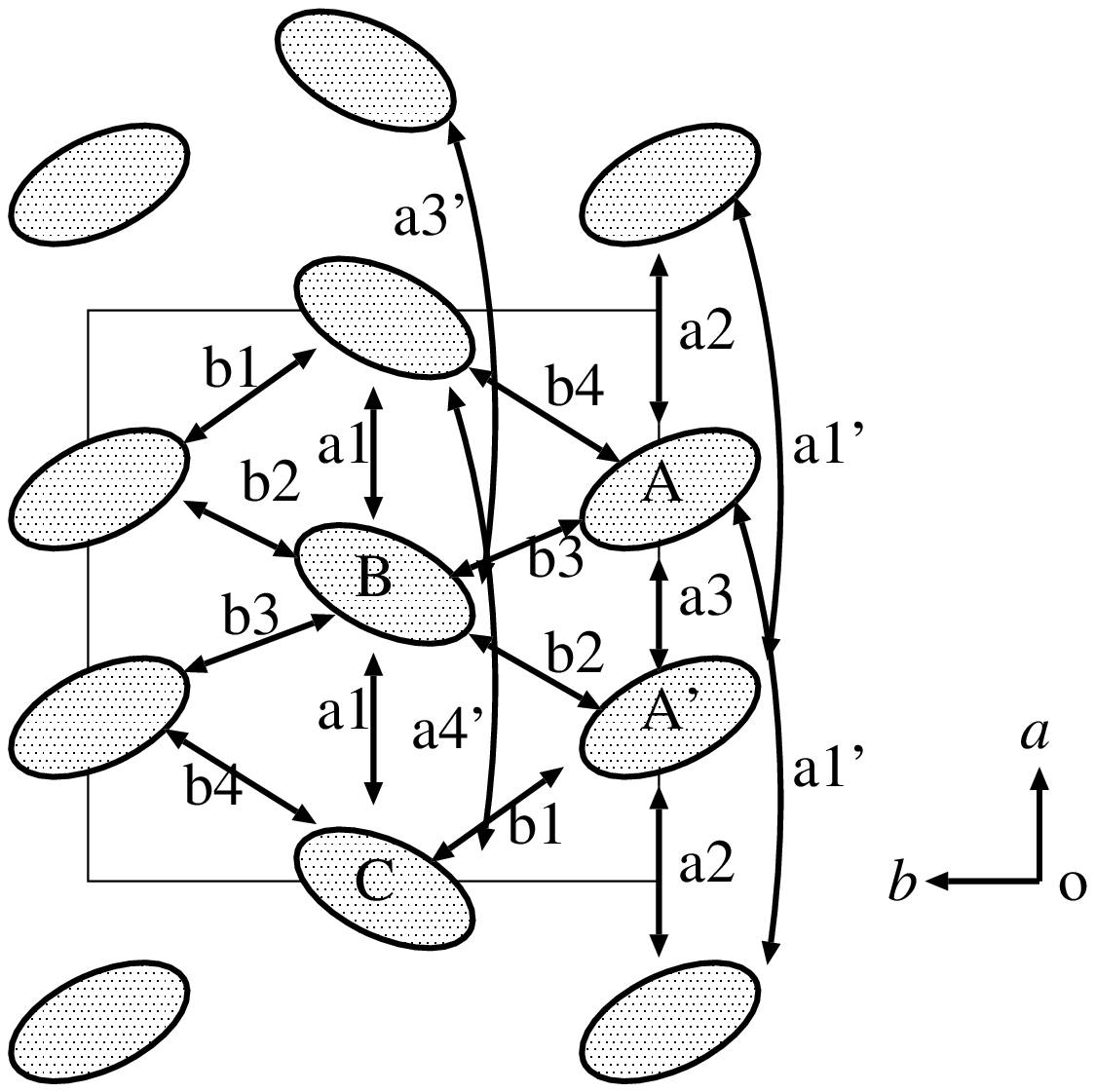}
\end{center}
\caption{
Crystal structures of conducting plane (the $ab$-plane) and transfer integrals 
 in Kondo's crystal axes.\cite{REF_KONDO_ATOMIC_PRIVATE,REF_HS_alphaET}
 Molecules A and A' are connected by inversion. 
a1', a3' and a4' are transfer integrals to the next nearest neighbor molecules along the $a$-axis.
}
\label{crystal}
\end{figure}

Next, we extract the parameters of the orthogonal tight-binding model to reproduce the first-principles band structure in the $ab$-plane.
We have found that the transfer integrals of the next nearest neighbor as well as those of the nearest neighbor are necessary to reproduce the electronic structure in detail. 
In this paper, we use not only the nearest neighbor parameters, a1, a2, a3, b1, b2, b3 and b4,  
 which have been introduced by Mori et al.,
but also a1', a3',and a4', which are the next nearest neighbor transfer integrals along the $a$-axis shown in Fig.~\ref{crystal}.\cite{REF_HS_alphaET}
The parameters are determined to reproduce not only the overall electronic structures, but also the detailed structure near the Dirac cone dispersion and near $(-\pi,\pi)$. 
We note that 
 including a1', a3' and a4' is not the only way to reproduce the band structure.
 Other sets of transfer integrals can also  reproduce the electronic structures. 
The Dirac cone dispersion itself can be reproduced without a1', a3' or a4'.


We show fitted parameters in Table~\ref{transferintegrals}.
We are not able to reproduce relative energies between $E(-\pi,0)$ and $E({\rm cone})$ at RT accurately.
However, the energy differences between the first-principles results and the tight-binding ones are a few of meV. Such a small difference is irrelevant because the thermal broadening of RT is much larger. The agreement is much better at 8~K.


Comparing the transfer integrals at ambient pressure and RT with those
of the semi-empirical estimation, the transfer integrals of b1-b4 are
found to agree quite well. 
On the other hand, 
it would be better to say that 
the transfer integrals along the $a$-axis cannot give good agreement with those in the semi-empirical estimation because the next nearest transfer integrals, a1',a3' and a4', are necessary. 
The dispersion, as well as the transfer integrals, along the $a^*$-axis
is much smaller than that along the $b^*$-axis as shown in
Fig.~\ref{dispersionAll}. In general, this may cause a problem in semi-empirical approximation when evaluating smaller values. 
Therefore, it is not surprising that the semi-empirical approximation
fails to yield precise estimates of transfer integrals along the $a$-axis.
These changes of transfer integrals may possibly affect the stability of the charge-ordering patterns.\cite{REF_SEO_REVIEW,REF_SEO_ALPHA,REF_TAKAHASHI_NMR,REF_KOBOYASHI_THEO_ANOTHER_SUPER} 

\begin{table}

\caption{
Fitted transfer integrals (eV) at ambient pressure (ap) and 
with uniaxial strain of 2~kbar along the $a$-axis (a2) and 
3~kbar along the $b$-axis (b3) at room temperature (RT) and 8~K.
a1-a4, b1-b4 and a1'-a3' are shown in Fig.~\ref{crystal}.
a1-a4 and b1-b4 under the tb column are the transfer integrals used in the semi-empirical approximation calculated 
as overlap integrals times $-10$~eV and are shown for comparison.\cite{REF_HS_alphaET}
}
\begin{tabular}{lrrrr} $T$=RT \\ \hline
   & ap & a2 & b3 & tb  \\ \hline
a1 & -0.0101 & -0.0173 & -0.0195 & -0.03  \\
a2 & -0.0476 & -0.0358 & -0.0393 & -0.049  \\
a3 & 0.0093 & 0.0029 & 0.0287 & 0.0018  \\
b1 & 0.1081 & 0.1075 & 0.1226 & 0.123  \\
b2 & 0.1109 & 0.1154 & 0.1176 & 0.142  \\
b3 & 0.0551 & 0.0559 & 0.0537 & 0.062  \\
b4 & 0.0151 & 0.0085 & 0.0156 & 0.023  \\
a1' & 0.0088 & 0.0085 & 0.0220  \\
a3' & 0.0019 & 0.0072 & -0.0151  \\
a4' & 0.0009 & -0.0012 & -0.0019 \\ \hline

\end{tabular}

\begin{tabular}{lrrr} $T$=8~K \\ \hline
        &ap &  a2  & b3 \\ \hline
a1 & -0.0267 & -0.0245 & -0.0267  \\ 
a2 & -0.0511 & -0.0461 & -0.0405  \\ 
a3 & 0.0323 & 0.0373 & 0.0397  \\ 
b1 & 0.1241 & 0.1276 & 0.1336  \\ 
b2 & 0.1296 & 0.1220 & 0.1251  \\ 
b3 & 0.0513 & 0.0507 & 0.0515  \\ 
b4 & 0.0152 & 0.0133 & 0.0190  \\ 
a1' & 0.0119 & 0.0247 & 0.0197  \\ 
a3' & 0.0046 & -0.0105 & -0.0118  \\ 
a4' & 0.0060 & -0.0019 & 0.0061  \\  \hline 

\end{tabular}\label{transferintegrals}
\end{table}

\begin{figure}
\begin{center}
\includegraphics[width=6.5cm]{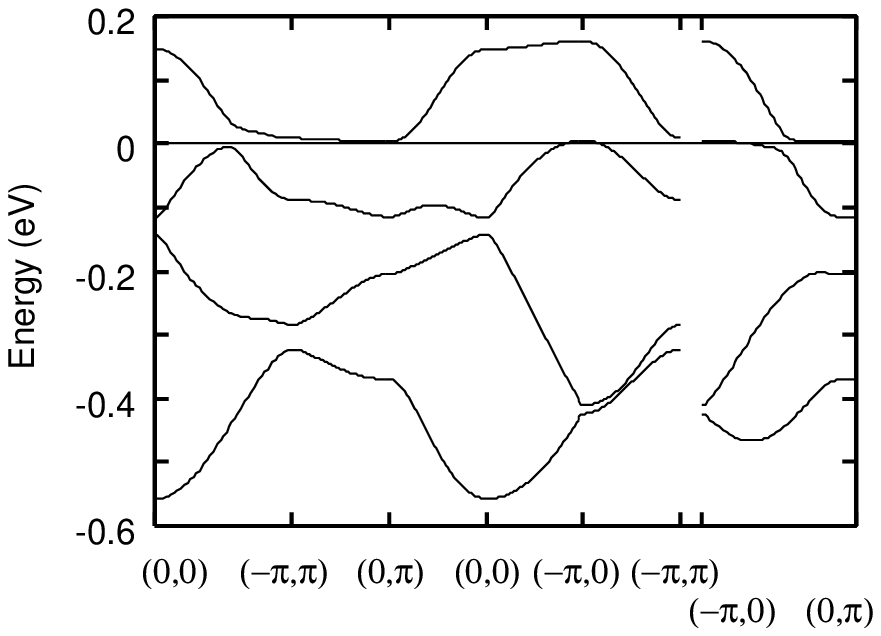}
\end{center}
\caption{Two-dimensional band structure in the semi-empirical approximation, which is  compared with Fig.~\ref{dispersionAll}.
The semi-empirical transfer and overlap integrals are from ref.~7.
Note that we do not use the fitted transfer integrals from Table~\ref{transferintegrals} in this figure.
We show this figure because the axes and $k$-points differ from study to study.
}
\label{hueckelband}
\end{figure}

Here we compare the first-principles band structure with the electronic structure in the semi-empirical parameters shown in Fig.~\ref{hueckelband}.\cite{REF_HS_alphaET}
The overall shapes agree quite well.
However, 
we find notable differences in the dispersion near the Fermi energy ($E=0$).
For example, let us consider the line from $(-\pi,\pi)$ to  $(0,\pi)$.
The bottom of the LUMO band lies at $(0,\pi)$  in the semi-empirical approximation, but it is at $(-\pi,\pi)$  in the first-principles result.
In the HOMO band, the top is located at $(-\pi,0)$ in the semi-empirical approximation, but  $E(-\pi/2,\pi/2)$  is higher than  $E(-\pi,0)$ in the first-principles result.
These changes make the Fermi surfaces different from each other. 
This difference in the $(-\pi,\pi)-(0,\pi)$ line in the semi-empirical approximation is related to the fact that the fitted tight-binding parameters along the $a$-axis are quite different from the H\"uckel ones. 

\begin{figure}
\begin{center}
\includegraphics[width=8.0cm]{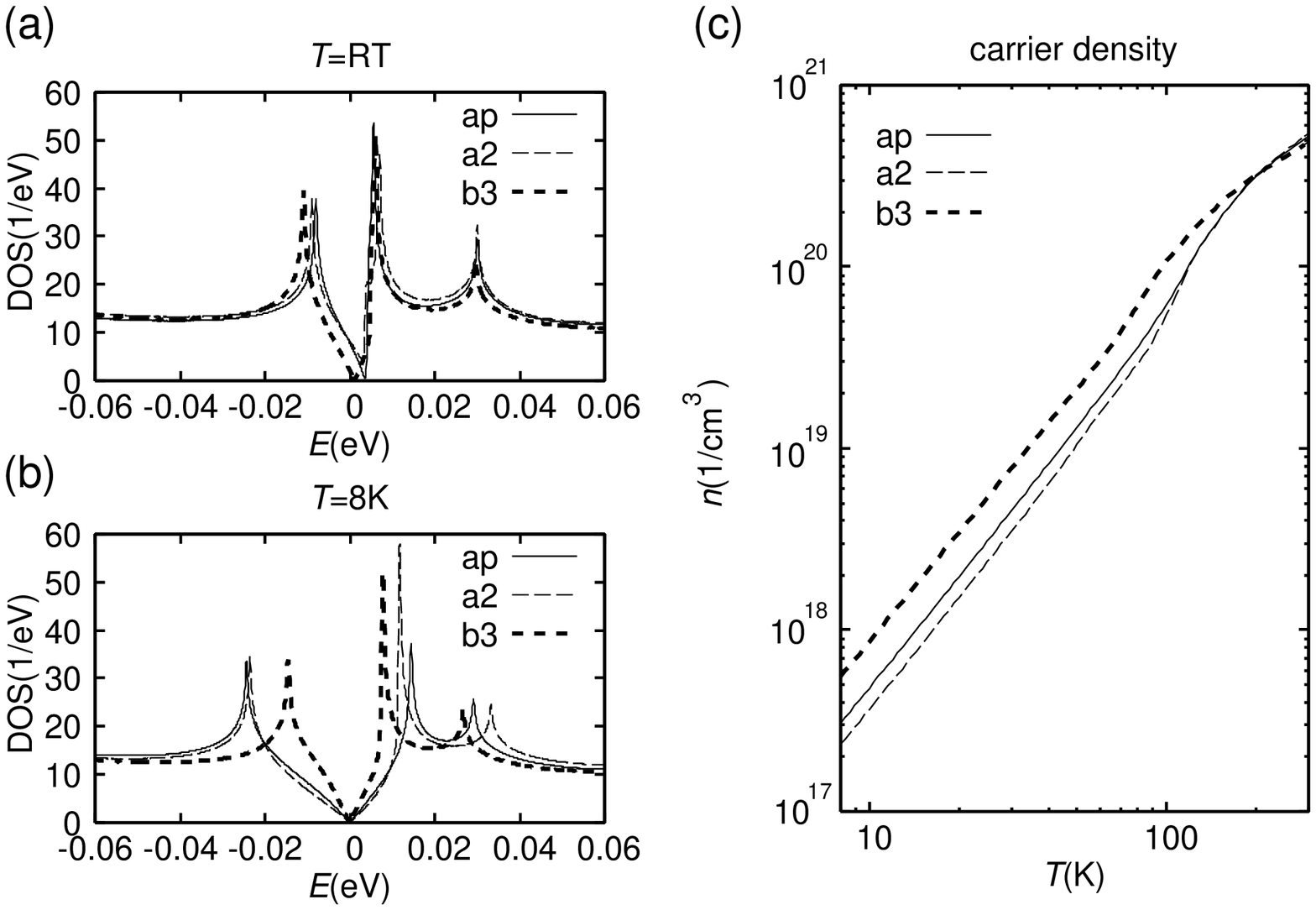} 
\end{center}
\caption{
The density of states at room temperature (RT) (a) and 8~K (b) at ambient pressure (ap, solid lines),
with uniaxial strain of 2~kbar along the $a$-axis (a2, thin dashed lines) and
 3~kbar along the $b$-axis (b3, thick dashed lines).
(c) Effective carrier density estimated as
$n(T)=\int_{-T}^{T} D(E,T) dE/V(T)$, 
where $D(E,T)$ is the density of states, and $V(T)$ is the cell volume at temperature $T$.
We interpolate 
values of the transfer integrals and the cell volume linearly as a function of temperature
 to calculate $D(E,T)$ and $V(T)$. 
}
\label{DOSfigure}
\end{figure}

The existence of the Dirac cone dispersion gives birth to interesting physical properties. We show the density of states at each strain and temperature in Fig.~\ref{DOSfigure}, using the fitted parameters and assuming the dispersion along the $c^*$-axis to be irrelevant.
$E=0$ corresponds to the chemical potential in the figure (this includes the effect of temperature broadening). 

First, let us examine the density of states at RT (Fig.~\ref{DOSfigure}~(a)). 
There exists a van Hove singularity quite near $E({\rm cone})$, which makes the density of state at $E=0$ finite due to the thermal broadening. Note that the thermal energy of RT corresponds to about 0.03~eV; therefore electronic structures of two peaks above and below $E=0$ contribute to the physical properties at RT.
The DOS with strain along the $b$-axis looks different from the others. 
The position of the van Hove singularity above the chemical potential with uniaxial strain of 3~kbar along the $b$-axis is higher than those at ambient pressure and with uniaxial strain of 2~kbar along the $a$-axis, as shown in Table.~\ref{Ediff}.

At 8~K (Fig.~\ref{DOSfigure}~(b)), 
 the linear behavior of the density of states near $E=0$ is clear. The nearest peak above the chemical potential ($E\sim +0.01 $~eV) corresponds to the van Hove singularity at $(-\pi,\pi)$ and the peak below the chemical potential ($E\sim -0.02 $~eV) corresponds to another van Hove singularity at $(-\pi,0)$. At this energy scale, we can see notable differences among the temperatures and uniaxial strains. While the density of states near $E=0$ with uniaxial strain of 2~kbar along the $a$-axis is almost the same as that at ambient pressure, the density of states with uniaxial strain of 3~kbar along the $b$-axis shows much larger values.

The property with strain along the $b$-axis looks different from the others at both temperatures.
The positions of van Hove singularities above the chemical potential  ($E(-\pi,\pi)-E({\rm cone})$ in the first-principles result)
 are summarized in Table~\ref{Ediff}.
The peak positions in the density of states between RT and 8~K are similar to Ishibashi et al.'s result\cite{REF_ISHIBASHI}. The difference is mainly due to the fact that we take into account of effects of the Dirac cone dispersion explicitly.

Next we calculate effective carrier density as a function of temperature. First, we calculate the density of states, $D(E,T)$, at each temperature using linearly interpolated transfer integrals at each strain; then we simply estimate the effective carrier densities as $n(T)=\int_{-T}^{T} D(E,T) dE/V(T)$, where $V(T)$ is the cell volume at temperature $T$. We show the result in Fig.~\ref{DOSfigure}~(c). At low temperatures, the carrier density is proportional to the square of the temperature, reflecting the linear density of states near the chemical potential shown in Fig.~\ref{DOSfigure}~(b). At about 100~K, there is a hump. This is due to the peak in the density of states above the chemical potential at $(-\pi,\pi)$. This hump would add an electron contribution to the Hall effect. The carrier density in our calculation is $4.8 \times 10^{17}$ cm$^{-3}$ at ambient pressure and $3.6 \times10^{17}$ cm$^{-3}$ at uniaxial strain of 2~kbar along the $a$-axis at 10~K.
 The experimental carrier density appears to be  proportional to the square of the temperature with uniaxial strain of 10~kbar along the $a$-axis, and its value is about $1 \times 10^{17}$ cm$^{-3}$ at 10~K. 
One cannot compare these results directly, because the value of  strain is greatly different.
However, their values agree quite well.

%
%

There is a discrepancy, however, in the sign of the Hall coefficient, $R_H$. 
Experimentally, the sign of $R_H$ is very sensitive. 
It depends on the value of hydrostatic pressure or uniaxial strain.
For example, with uniaxial strain of 10~kbar along the $a$-axis, $R_H$ is negative at low temperatures, while it is positive at high temperatures. 
On the other hand, with high hydrostatic pressure (20~kbar), it is positive over the whole temperature range.
\cite{REF_TAJIMA_SUPER, REF_KAJITA_MOBILITY}

If the particle-hole symmetry is satisfied, $R_H=0$. 
Therefore, asymmetry is necessary to give nonzero $R_H$. 
In the calculated results, the nearest peak from the chemical potential is located on the positive energy side at ($-\pi$,$\pi$) at 8~K. 
For example,  at ambient pressure or with uniaxial strain along the $a$-axis, $E(-\pi,\pi)-E({\rm cone})=8\sim10$~meV at 8~K.
The second nearest peak from the chemical potential is on the negative energy side at ($-\pi$, 0). 
However, $E({\rm cone})-E(-\pi,0)$ is twice as large as $E(-\pi,\pi)-E({\rm cone})$ at 8~K.
This makes the chemical potential slightly lower than $E({\rm cone})$, which results in a positive value of $R_H$. 
In contrast, $R_H$ becomes negative if 
$E(-\pi,0)$  ($<E({\rm cone})$)
is closer to $E({\rm cone})$ than
$E(-\pi,\pi)$ ($>E({\rm cone})$) is.

Provided that the linear density of states of the Dirac cone dispersion governs the temperature dependence of the carrier density with uniaxial strain of 10~kbar along the $a$-axis, 
a possible origin of the change of the sign of $R_H$ is that both the peak at 
$(-\pi,\pi)$ on the positive energy side and the peak at $(-\pi,0)$ on the negative energy side would be easily shifted due to strain. 
Considering that 10~kbar is much larger than 2~kbar, these differences could arise. 
We do not discuss the case of the hydrostatic pressure of 20~kbar, because the value is too different from strain of 2 or 3~kbar calculated in this study. 
We note that $E(-\pi,0)$ ($<E({\rm cone})$) is much closer to the $E({\rm cone})$ than $E(-\pi,\pi)$ ($>E({\rm cone})$) without the structural relaxation.

The experiment also shows that $R_H>0$ at higher temperatures. 
However, it is difficult to conclude which contribution is dominant at RT because the dispersion is very complex. 
The experimental crystal structures with high hydrostatic pressure and with high uniaxial strain are unknown. 
Further theoretical and experimental studies  on this problem are necessary.

In this paper, we do not discuss the origin, or the condition of the existence, of the anisotropic Dirac cone dispersion at the Fermi level and {\it at a general $k$-point}.
Here, we mention that
these first-principles calculations and their tight-binding fitting show
that the situation is expected to be realized in a rather wide range of parameters (temperature and strain).
The condition of the existence in simple tight-binding models can be found in ref.~10.


\subsection{ Electronic structure along the c-axis}
\label{sec_interlayer}

Next, we consider the dispersion along the $c^*$-axis, which is thought to be very small and has so far been neglected in the semi-empirical parameterization and tight-binding models. However, there is a possibility that the transfer integrals along the c-axis are relevant at low temperatures because of the semimetallic nature. 

First let us examine the overall nature. 
Table~\ref{Cdiff} shows the energy differences, $E(\kz=0)-E(\kz=\pi)$, along the $c^*$-axis. $(\kx,\ky)$=($-\pi/2$,$\pi/2$) is near the crossing point of the Dirac cone dispersion. 
The band dispersion along the $c^*$-axis is 4~meV at the greatest. 
The band dispersion of the HOMO band at $(-\pi,0)$ is very small. 
These reflect the fact that the nearest molecule perpendicular to the $ab$-plane is along the (1/2,1/2,1) direction.
These values are clearly much smaller than those of the band dispersion in the $a^*b^*$-plane.

It has been thought that this material is a two-dimensional conductor because the resistivity along the $c$-axis is $10^2$ times larger than those in the $ab$-plane (at ambient pressure).\cite{REF_FIRST_ALPHA}
Whether the transfer integrals along the $c^*$-axis ($\tz$) are relevant or not is an interesting problem. If carriers are confined in the conducting layer, or conductance along the $c$-axis does not occur via coherent hopping, then it will be safe to neglect $\tz$ in the analysis of transport properties within the $ab$-plane. However, $\tz$ may be relevant at temperatures lower than the energy scale of $\tz\sim 10$~K; then we must take into account of the effect of $\tz$. If the coherent hopping occurs between BEDT-TTF molecules directly along the $c$-axis, more calculations are necessary. 

\begin{figure}
\begin{center}
\includegraphics[width=7cm]{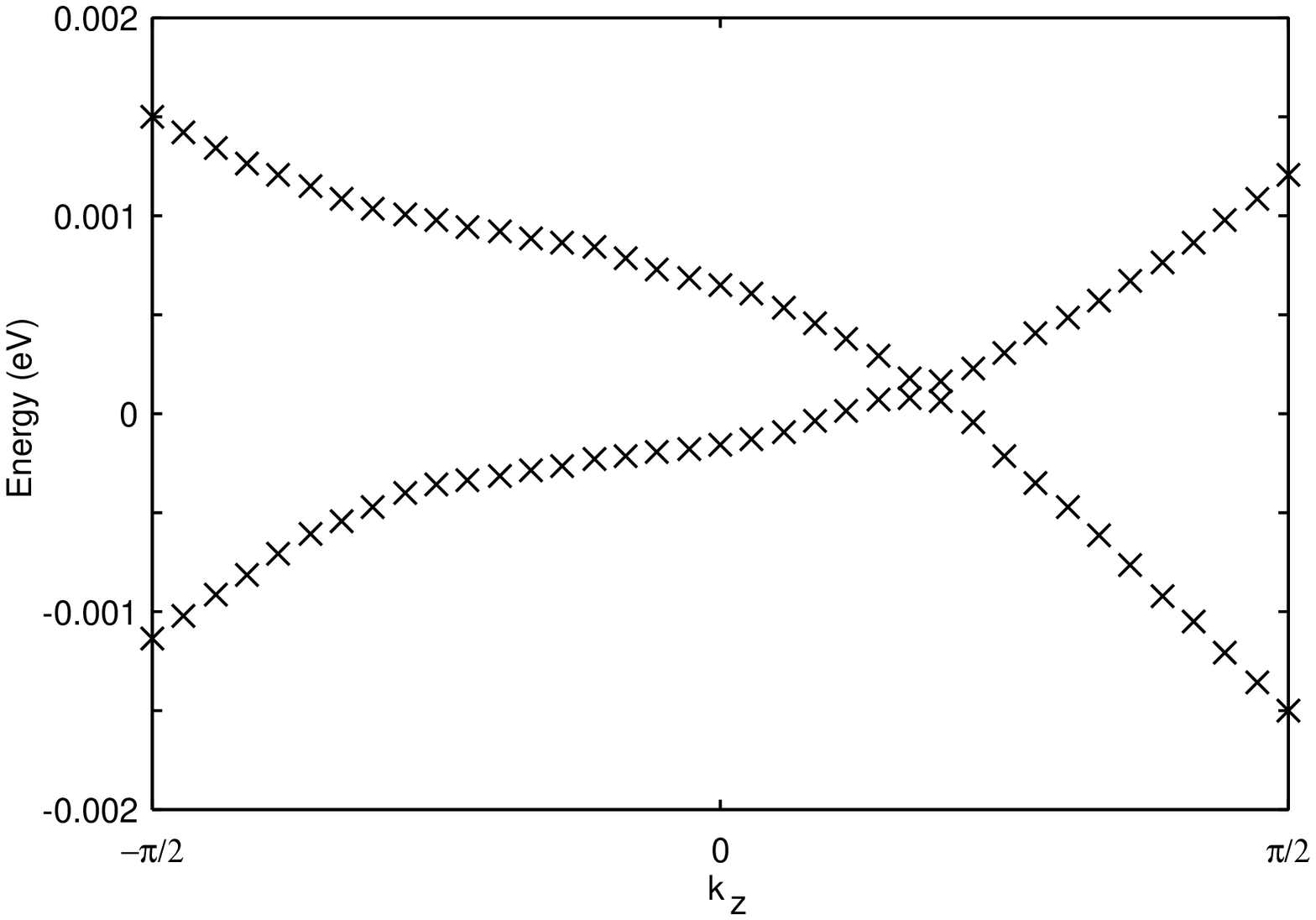}
\end{center}

\caption{
Electronic structure along $c^*$-axis near $(\kx,\ky)$ of the crossing point of 
the Dirac cone dispersion.
We take the gapless point to be at $E=0$.
}
\label{dispCaxis}
\end{figure}

We briefly look at the dispersion near the Dirac cone dispersion. 
We show the dispersion along the $c^*$-axis at some $k$-point near the crossing point of the Dirac cone dispersion in Fig.~\ref{dispCaxis}. 
It is not easy to predict the exact $k$-point where the gap is zero because this is a numerical study.
The figure shows that a gapless point exists at $\kz\sim0.3\pi$ when the energy gap at $\kz=0$ is not zero.
We expect that the gapless $\kz$-point depends on $(\kx,\ky)$,
and that a gapless point exists at $\kz=0$ at some $(\kx,\ky)$ near the $k$-point shown in the figure.
In order to understand this behavior completely, we must perform many more expensive calculations. 
Therefore, we leave further study of this problem for future studies. 

Finally, we note the possible Fermi surfaces. 
There are two kinds of possible Fermi surfaces (lines) at low temperatures if there are crossing points at some $(\kx,\ky)$, depending on the value of $\kz$. One possibility is Fermi lines stretching along the $c^*$-axis. 
However, it is not natural for the Fermi level to always be at the crossing points.
The other possibility is more probable: acicular Fermi surfaces pointing along the $c$-axis. The Fermi velocity along the $c^*$-axis is much smaller than those along the $ab$-plane. 
Note that the density of states at $|E| \simle 1$~meV$\sim$10~K may be changed due to this three-dimensionality if the dispersion along the $c^*$-axis is relevant.

\begin{table}

\caption{
Energy difference (meV) along the $c^*$-axis at ambient pressure (ap), with uniaxial strain of 2~kbar along the $a$-axis (a2) and  3~kbar along the $b$-axis (b3), at 8~K and room temperature (RT).
Values show $E(k_\perp,\kz=0)-E(k_\perp,\kz=\pi)$ at each $k_\perp=(\kx,\ky)$. }

\begin{tabular}{lccc}
$T$=8~K \\ \hline
width & ap & a2 & b3 \\ 
\hline $(-\pi,\pi)$ \\
LUMO & 2.6 & 2.6 & 2.5 \\
HOMO & -3.8 &-4.0 & -3.7 \\
\\
$(-\pi/2,\pi/2)$ \\
LUMO & -2.0 & -2.1 & -2.0  \\
HOMO & 3.0 & 3.2 & 3.1 \\
\\
$(-\pi,0)$ \\
LUMO & -1.2 & -2.3 & -2.3 \\
HOMO & -0.2 & -0.2 & -0.7\\ \hline
\end{tabular}

\begin{tabular}{lccc}
$T$=RT \\ \hline
width & ap & a2 & b3 \\ \hline
$(-\pi,\pi)$ \\
LUMO & 2.3 & 2.3 & 2.1 \\
HOMO & -3.8 & -3.7 & -3.8 \\
\\
$(-\pi/2,\pi/2 )$\\
LUMO & -2.3 & -2.4 & -2.4 \\
HOMO & 2.7 & 2.7 & 2.8 \\
\\
$(-\pi,0)$ \\
LUMO & -2.2 & -2.3 & -2.6 \\
HOMO & 0.2  & 0.2 & 0.1 \\ \hline
\end{tabular}

\label{Cdiff}
\end{table}

\section{Summary}
We have relaxed atomic positions of $\alpha$-(BEDT-TTF)$_2$I$_3$ at ambient pressure and with uniaxial strain along the $a$- and $b$-axes to determine the electronic structure by the first-principles calculation.
We have confirmed the existence of anisotropic Dirac cone dispersion in all the cases. 
Uniaxial strain of 3~kbar along the $b$-axis slightly increases the bandwidths, while uniaxial strain of 2~kbar along the $a$-axis decreases or does not change the bandwidths, though the overall effects of pressure to the bandwidths are not large. 
The effects of thermal contraction on the electronic structure are much stronger than those of uniaxial strain. 
There are two main effects of temperature.
One is the increase of the bandwidths with decreasing temperature.
The other is the change of the electronic structure near the chemical potential.
Only the Dirac cone dispersion appears near the chemical potential in the energy range of the temperature at 8~K, while the electron pocket at $(-\pi,\pi)$ and then the hole pocket at $(-\pi,0)$ appear within the energy range of the temperature, as well as the Dirac cone dispersion, at RT.
This Dirac cone dispersion yields the linear density of states, which gives the $T^2$ dependence of the carrier density up to about 100~K. 
If the effects of uniaxial strain of 10~kbar along the $a$-axis are almost the same as those of 2~kbar strain, the theoretical carrier density is comparable to the experimental value. The effects of uniaxial strain are different between the $a$-axis and the $b$-axis. uniaxial strain along the $a$-axis gives almost the same density of states as that obtained at ambient pressure, at least in the energy range of temperature, while strain along the $b$-axis gives a much large value of the density of states at 8~K. It is apparent that the effects of uniaxial strain are different between the $a$- and $b$- axes on a low-energy scale. These theoretical findings may have some connections with the experimental phase diagram that depends on the direction of strain.


\section*{Acknowledgements}
We thank  with Dr. S. Ishibashi, Dr. M. Onoda, Dr. A. Kobayashi, Professor Y. Suzumura, Dr. R. Kondo, Professor S. Kagoshima, Dr. N. Tajima, Professor Y. Nishio and Professor K. Kajita useful discussion. 
 This work is partially supported by a Grant-in-Aid for Scientific Research on Priority Areas of Molecular Conductors (No. 16038227) from MEXT of the Japanese Government.



\begin{thebibliography}{99}



\bibitem{REF_KAJITA_MOBILITY} K. Kajita, N. Tajima, A. Ebina-Tajima and Y. Nishio: Synth. Met. {\bf 133-134} (2003) 95.

\bibitem{REF_TAJIMA_SUPER} N. Tajima, A. Ebina-Tajima, M. Tamura, Y. Nishio and K. Kajita: J. Phys. Soc. Jpn. {\bf 71} (2002) 1832.

\bibitem{REF_KONDO_STRAIN} R. Kondo and S. Kagoshima: J. Phys. IV France, {\bf 114} (2004) 523.


\bibitem{REF_KAGOSHIMA_REV} S. Kagoshima and R. Kondo: Chem. Rev. {\bf 104} (2004) 5593.



\bibitem{REF_KONDO_ATOMIC_PRIVATE} R. Kondo and S. Kagoshima: private communications.


\bibitem{REF_HS} T. Mori, A. Kobayashi, Y. Sasaki, H. Kobayashi, G. Saito and H. Inokuchi: Bull. Chem. Soc. Jpn. {\bf 57} (1984) 627.

\bibitem{REF_HS_alphaET} T. Mori, A.Kobayashi, T. Sasaki, H. Kobayashi, G. Saito and H. Inokuchi: Chem. Lett. {\bf 1984} (1984) 957.

\bibitem{REF_KOBAYASHI_THEO_SUPER} A. Kobayashi, S. Katayama, K. Noguchi and Y. Suzumura: J. Phys. Soc. Jpn. {\bf 73} (2004) 3135.

\bibitem{REF_KOBOYASHI_THEO_ANOTHER_SUPER} A. Kobayashi, S. Katayama and Y. Suzumura:  J. Phys. Soc. Jpn. {\bf 74} (2005).

\bibitem{REF_KATAYAMA_DIRAC} S. Katayama, A. Kobayashi and Y. Suzumura: to be published in J. Phys. Soc. Jpn.


\bibitem{REF_MIYAZAKI_BEDT} T. Miyazaki and H. Kino: Phys. Rev. B {\bf 68} (2003) 220511.

\bibitem{REF_MIYAZAKI_BEDT2} T. Miyazaki and H. Kino: to be appeared in Phys. Rev. B.

\bibitem{Xraysymmetry} We assume P$\bar{1}$ symmetry in all the cases. 

\bibitem{REF_HERRING} C. Herring: Phys. Rev. {\bf 52} (1937) 365.

\bibitem{REF_SEO_REVIEW}  H. Seo, C. Hotta and H. Fukuyama: Chem. Rev. {\bf 104} (2004) 5005.

\bibitem{REF_SEO_ALPHA} H. Seo: J. Phys. Soc. Jpn. {\bf 69} (2000) 805.




\bibitem{REF_TAKAHASHI_NMR} T. Takahashi: Synth. Met. {\bf 133-134} (2003) 261.


\bibitem{REF_ISHIBASHI} S. Ishibashi, T. Tamura, M. Kohyama and K. Terakura: to be appeared in J. Phys. Soc. Jpn. (2006, No.~1)

\bibitem{REF_FIRST_ALPHA} K.I. Pohkodnya, Y.V. Sushko and M.A. Tanatar: Sov. Phys. JETP {\bf 65}  (1987) 795.


\end{thebibliography}
\end{document}